\documentclass[twocolumn]{aastex631}

\usepackage{enumitem}
\usepackage{hyperref}
\usepackage{multirow}
\usepackage{soul}

\shorttitle{\ion{H}{1} in M94 with FAST}
\shortauthors{Zhou et al.}

\graphicspath{{./}{figures/}}

\begin{document}

\title{FAST reveals new evidence for M94 as a merger}

\author{Ruilei Zhou}
\affiliation{National Astronomical Observatories, Chinese Academy of Sciences, A20 Datun Road, Beijing 100101, China}
\affiliation{University of Chinese Academy of Sciences, Beijing 100049, China}
\affiliation{CAS Key Laboratory of FAST, National FAST, National Astronomical Observatories, \\
Chinese Academy of Sciences, Beijing 100101, China}

\correspondingauthor{Ming Zhu}

\author[0000-0001-6083-956X]{Ming Zhu}
\affiliation{National Astronomical Observatories, Chinese Academy of Sciences, A20 Datun Road, Beijing 100101, China}
\affiliation{University of Chinese Academy of Sciences, Beijing 100049, China}
\affiliation{CAS Key Laboratory of FAST, National FAST, National Astronomical Observatories, \\
Chinese Academy of Sciences, Beijing 100101, China}
\affiliation{Guizhou Radio Astronomical Observatory, Guizhou University, Guiyang 550000, People's Republic of China}
\email{mz@nao.cas.cn}

\author[0000-0001-7949-3407]{Yanbin Yang}
\affiliation{GEPI, Observatoire de Paris, CNRS, Place Jules Janssen 92195, Meudon, France}

\author{Haiyang Yu}
\affiliation{National Astronomical Observatories, Chinese Academy of Sciences, A20 Datun Road, Beijing 100101, China}
\affiliation{University of Chinese Academy of Sciences, Beijing 100049, China}
\affiliation{CAS Key Laboratory of FAST, National FAST, National Astronomical Observatories, \\
Chinese Academy of Sciences, Beijing 100101, China}

\author[0000-0003-0804-9055]{Lixia Yuan}
\affiliation{Purple Mountain Observatory and Key Laboratory of Radio Astronomy, Chinese Academy of Sciences, 10 Yuanhua Road, Qixia District, Nanjing 210033, China}

\author{Peng Jiang}
\affiliation{National Astronomical Observatories, Chinese Academy of Sciences, A20 Datun Road, Beijing 100101, China}
\affiliation{CAS Key Laboratory of FAST, National FAST, National Astronomical Observatories, \\
Chinese Academy of Sciences, Beijing 100101, China}
\affiliation{Guizhou Radio Astronomical Observatory, Guizhou University, Guiyang 550000, People's Republic of China}

\author{Wenzhe Xi}
\affiliation{University of Chinese Academy of Sciences, Beijing 100049, China}
\affiliation{Yunnan Observatories, Chinese Academy of Sciences, Kunming 650011, China}

\begin{abstract}

We report the first high-sensitivity \ion{H}{1} observation toward the spiral galaxy M94 with the Five-hundred-meter Aperture Spherical radio Telescope (FAST). From these observations, we discovered that M94 has a very extended \ion{H}{1} disk, twice larger than that observed by THINGS, which is accompanied by an \ion{H}{1} filament and seven HVCs  (high velocity clouds) at different distances. The projected distances of these clouds and filament are less than 50~kpc from the galactic center. We measured a total integrated flux (including all clouds/filament) of 127.3 ($\pm$1) Jy km s$^{-1}$, corresponding to a \ion{H}{1} mass of (6.51$\pm$0.06)$\times10^{8} M_{\odot}$ , which is 63.0\% more than that observed by THINGS. By comparing numerical simulations with the \ion{H}{1} maps and the optical morphology of M94, we suggest that M94 is likely a remnant of a major merger of two galaxies, and the HVCs and \ion{H}{1} filament could be the tidal features originated from the first collision of the merger happened about 5~Gyr ago.  
Furthermore, we found a seemingly isolated  \ion{H}{1} cloud at a projection distance of 109~kpc without any optical counterpart detected. We discussed the possibilities of the origin of this cloud, such as dark dwarf galaxy and RELHIC (REionization-Limited \ion{H}{1} Cloud). Our results demonstrate that high-sensitivity and wide-field \ion{H}{1}  imaging is important in revealing the diffuse cold gas structures and tidal debris which is crucial to understanding the dynamical evolution of galaxies.

\end{abstract}


\section{Introduction}

NGC 4736 ($=$M94) is a nearby bright galaxy at a distance of 4.66 Mpc \citep{K+05}, which was classified as a (R)SA(r)ab-type galaxy \citep{de+91} and reclassified as (R)SAB(rs)ab-type by \citet{B+07}. 
It is the brightest galaxy of the Canes Venatici I cloud (CVnI) near the center of the CVnI cloud and it is the main disturber relative only to 10 neighboring galaxies  \citep{K+05}. M94 is the closest example of a LINER 2 nucleus, although whether the LINER 2 is powered by a low-luminosity active galactic nucleus (LLAGN) is uncertain \citep{R+01}. 

Multi-wavelength images of M94 have been observed to study its morphology, such as optical (SDSS), UV \citep{M+95}, far-infrared \citep{S+91}, near-infrared \citep[2MASS,][]{J+03}, and 21 cm \ion{H}{1} line \citep[THINGS,][]{de+08}. According to \citet{T+09}, NGC 4736 can be divided into five main regions: a) a bulge ($R\textless15\arcsec$), b) an inner ring ($R\approx 45\arcsec$), c) an outer spiral structure with oval stellar distribution ($R\approx 220\arcsec$), d) a zone of low surface brightness, e) a faint outer ring ($R\approx 350\arcsec$). 
However, non-optical data suggest that there is a disk with a spiral arm structure rather than a stellar ring in the outer region of NGC 4736, such as \ion{H}{1} (\citealt{W+08}, their fig.~51; \citealt{de+08}, their fig.~80; \citealt{B+95}, his fig.8), UV (\citealt{T+09}, their fig. 2), and infrared (\citealt{T+09}, their fig. 2). M94 is a post-starburst galaxy: In FUV, \citet{K+93} found that the off-nuclear continuum is flat with little trace of the emission-like features around 1900 $\rm{\mathring{A}}$ seen in the nuclear spectrum, and \citet{W+01} thought it is consistent with an aging burst which is spatially more extended than the centrally-concentrated stars of the older disk and bulge; \citet{M+04} found evidence for a fossil starburst nucleus, which is a kinematically detached innermost area defined by a clear dip in the stellar velocity dispersion with a size about 150~pc. \citet{S+98} observed the post-starburst nucleus of M94 in ${}^{12}$CO (J=1-0), and they thought bar-driven instability fueling or minor-merger driven fueling (or triggering) may be the possible scenarios for the nuclear starburst.

Although many studies show that M94 is a rather isolated galaxy \citep{Wa+16}, there are much evidence that M94 is related to interaction or merger. As a post-starburst galaxy, \citet{M+05} noted a second UV source which is offset from the galaxy nucleus by 2$\arcsec$.5 ($\sim$60~pc in projection), and they thought that the off-nuclear could be the active nucleus of a galaxy that had merged with M94. They also suggested that this merger may trigger the past starburst and the peculiar morphological and kinematic features observed in M94. \citet{C+11} drew a conclusion that M94 may be in the final stage of a merger by comparing the intricate structure of off-nuclear compact source detections in X-ray, radio, and UV. \citet{Ko+05} reported another possibility for the off-nuclear source that could be the result of jet activities in the nucleus. The kinematics of planetary nebulae (PNe) also show evidence of flaring in the old stellar population \citep{H+09,HC+09}, which may indicate some past perturbation. M94 has a bright anti-truncated outer disk \citep{T+09} and its rotation curves  decline by $\sim$30\% from the inner region ($\sim$3~kpc) to the outer region ($\sim$9~kpc) \citep{M+95,de+08}. In the simulations of \citet{P+06}, extended disks may be the result of a tidal disrupted dwarf galaxy coplanar with the disk of the host galaxy in prograde orbit, and the kinematic average rotational motion of the stars forming the extended disk is more than 30-50~km~s$^{-1}$ lower than the host’s circular velocity. Thus the much declining rotation curve could favor a merger origin for the outer disk of M94. However, if all the outer disk were entirely the result of a disrupted dwarf, M94 would have a high merger ratio 1:4 \citep{T+09}, so we could not ignore the possibility that the accretion destroy the inner disk \citep{Ho+09}. And in the merging simulations of massive satellites with disk galaxies \citep{V+99}, the tilt for the outer regions is much smaller than that observed \citep{T+09}. \citet{T+09} also found that the specific star formation rate of the outer disk is twice that of the inner disk, and they listed two origins of it: merger and an oval distortion in the inner disk. Using numerical simulations, they found the latter origin provided a good match to the observed structure of M94. Thus the existing observational evidences are not enough to prove that M94 has undergone a merger event. To further investigate the evolution history of M94, more high sensitivity observations are needed. 

In this paper, we present a wide field \ion{H}{1} imaging study of M94 with the Five-hundred-meter Aperture Spherical radio Telescope (FAST). Our newly discovered morphological and kinematic features could be used to constrain models for galaxy interactions and evolution. In Section 2, we describe the details of the observations and data reduction. In Section 3, we describe the properties of these \ion{H}{1} features in detail and discuss the possible origins of the discoveries in Section 4. Finally, we summarize our results in Section 5. All velocities quoted in the paper are heliocentric, unless otherwise specified.

\section{Observation and data reduction} 

\begin{figure*}
\begin{center}
	\includegraphics[width=0.7\textwidth]{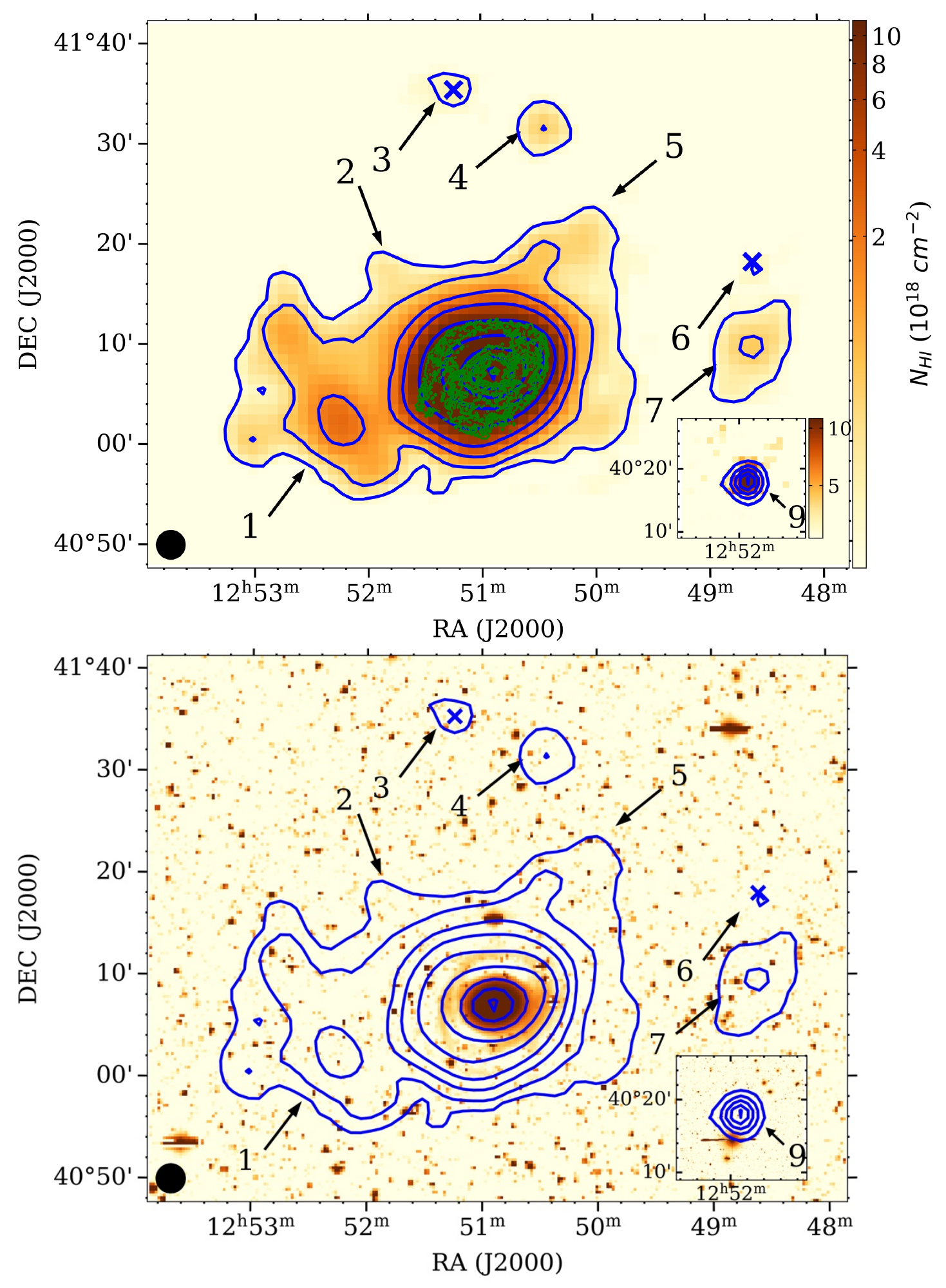}
    \caption{
    {\it Top}: The \ion{H}{1} integrated intensity map of M94 region with the THINGS and FAST contours overlaid on it. The blue contours are FAST data integrated over 114-504~km~s$^{-1}$ and contour levels are 0.2, 0.7, 1.9, 4.4, 11.0, 21.0, 33.4, 45.8, 58.3$\times10^{19}$~cm$^{-2}$. Green ones are THINGS image integrated over 144-490~km~s$^{-1}$, and contour levels start at 1.3$\times10^{20}$~cm$^{-2}$, and increments by 2.5$\times10^{20}$~cm$^{-2}$. {\it Bottom}: the same FAST contours overlaid on a DESI-LS optical image. The lower right corner of the both panels shows Cloud 9 integrated over 290-319~km~s$^{-1}$ with contour levels start at $6.7\times10^{17}$~cm$^{-2}$ in steps of $6.7\times10^{17}$~cm$^{-2}$. The numbers represent names of the \ion{H}{1} features indicated by the corresponding arrows. The crosses stand for possible features of HVC 3 and 6.
    }
\label{fig:m94_mom0}
\end{center}
\end{figure*}

Using the FAST (\citealt{N+11}; \citealt{J+19}; \citealt{J+20}), we mapped a sky region of Right Ascension of $191.57^{\circ}\textless\alpha\textless193.84^{\circ}$, Declination of $40.03^{\circ}\textless\delta\textless41.82^{\circ}$ and velocity range of 114.2-504.2~km~s$^{-1}$. FAST is located in Guizhou, China, with 500~m aperture and 300~m illuminated aperture. The observation was carried out with the 19-beam receiver which has a frequency range from 1050~MHz to 1450~MHz. We select Spec(W) spectrometer which has a bandwidth of 500MHz and 655536 channels, corresponding to a spectral resolution of 1.67~km~s$^{-1}$. With a $23.4^{\circ}$ rotation of the 19-beam receiver platform, the beam tracks are equally spaced in Declination with 1$\arcmin$.14 spacing. The half-power beamwidth (HPBW) was about 2$\arcmin$.9 at 1.4 GHz for each beam and the pointing accuracy of FAST was about 12$\arcsec$.

The data is composed of two parts. Part I was observed on 2021 December 13,15,16,23,25,26, carried out with the drift scan mode. Part II was observed on 2021 May 01, and carried out with the Multibeam on-the-fly (OTF) mode, which focused on the region around M94 in order to improve the sensitivity to detect weak features outside the M94 disk.   We performed flux calibration by injecting a 10K calibration signal (CAL) every 32 s for a duration of 1 s to calibrate the antenna temperature. The data were reduced using  HIFAST (Y. Jing et al. 2023, in preparation) data reduction pipeline. And we made baseline correction with the asymmetrically reweighted penalized least-squares algorithm (arsPLs, \citealt{B+15}). After the spectra were fully calibrated, we gridded them into an image with 1$\arcmin$ spacing, and created the data cube in the standard FITS format. More detailed procedures are described in \citet{X+21}. The data were then smoothed from 1.67~km~s$^{-1}$ velocity resolution to 4.88~km~s$^{-1}$. Since we used the same techniques as HIPASS to calibrate and grid the data, we made flux corrections according to \citet{B+01}. The RMS brightness temperature sensitivity of Part I is approximately 9.3 mK or 0.6 mJy beam$^{-1}$ per channel, while that of the region with both Part I and II data combined  is 6.2 mK or 0.4 mJy beam$^{-1}$ per channel. In the following study, the $\sigma$ refers specifically to the $\sigma$ of Part II, since most features were studied through Part II.

\begin{figure*}
\begin{center}
	\includegraphics[width=0.7\textwidth]{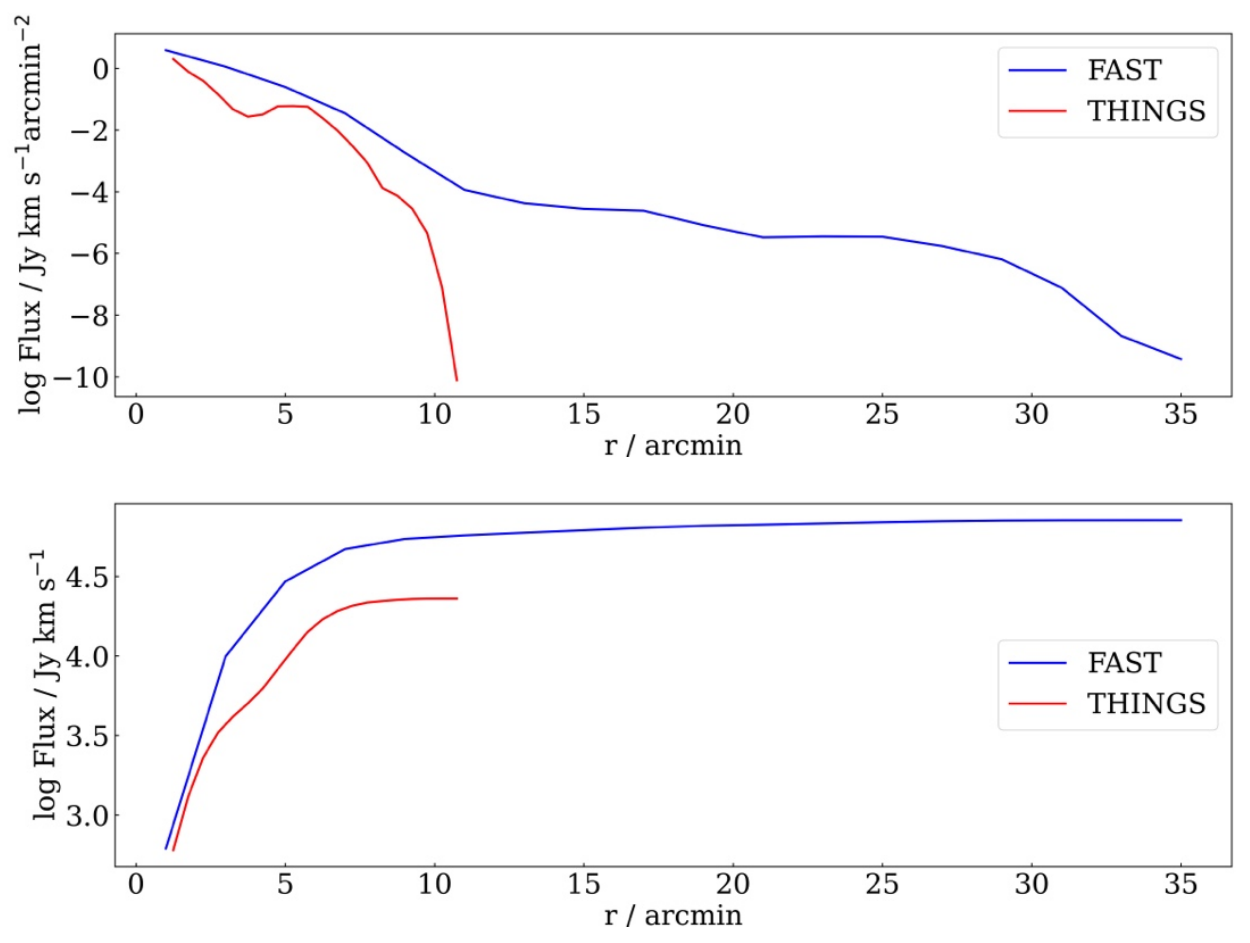}
    \caption{
    {\it Top}: The azimuth averaged flux profiles of FAST (blue curve) and THINGS map (red curve) which are varied with the distance from the center of M94. {\it Bottom}: the accumulated profile of flux of FAST (blue curve) and THINGS map (red curve) varied with the distance from the center of M94.
    }
\label{fig:m94_profile}
\end{center}
\end{figure*}

\section{Results} 

\subsection{Extended \ion{H}{1} Map}

The top panel of Fig.~\ref{fig:m94_mom0} shows the \ion{H}{1} column density distribution of M94 integrated over 114-504~km~s$^{-1}$ overlaid with the \ion{H}{1}  integrated flux density contours detected by the VLA (green) from THINGS \citep{de+08,W+08}. In this figure, the lower right subpanel is integrated over 290-319~km~s$^{-1}$ to show a weak \ion{H}{1} cloud far away from the central region. The bottom panel shows the same FAST detected contours superimposed on the  optical image of M94 from the DESI Legacy Imaging Surveys \citep[DESI-LS;][]{D+19}. The THINGS survey has a high spatial resolution  ($\sim6\arcsec$)  which can reveal the detail \ion{H}{1} distribution of the galaxy disk, while the FAST data can better show the diffuse \ion{H}{1} gas outside the disk. The FAST contour levels starting at 3~$\sigma$ ($2.5\times10^{18}$~cm$^{-2}$ for M94 region, and $6.7\times10^{17}$~cm$^{-2}$ for the lower right subpanel) is much lower than that of THINGS map which is $1.3\times10^{20}$~cm$^{-2}$. The moment-0 map was made using the spectral-cube algorithm \footnote{https://spectral-cube.readthedocs.io/en/latest/moments.html
\#moment-maps} and smoothed by a Guassian filter (kernel=3$\times$3). A few obvious bad pixels were masked manually. From Fig.~\ref{fig:m94_mom0}, we can see that the FAST map reveals a large amount of diffuse \ion{H}{1} gas outside the \ion{H}{1} disk covered in the THINGS map. The THINGS map covers the central  $\sim$ 140~arcmin$^{2}$ area, while the FAST's map covers three times more extended area than that of THINGS, and allows us to detect huge amount of low column density  \ion{H}{1} gas. 

\subsection{Total \ion{H}{1} Flux}

By integrating all the \ion{H}{1} flux in Fig.~\ref{fig:m94_mom0}, including the \ion{H}{1} structures discussed in the next section, we measured a total integrated flux of 127.3($\pm$1)Jy~km~s$^{-1}$, or \ion{H}{1} mass of (6.51$\pm$0.06)$\times10^{8} M_{\odot}$ for M94. Within the region covered by THINGS, the integrated flux observed by FAST is 97.0~Jy~km~s$^{-1}$, while that from THINGS measurement \citep{W+08} is 78.1~Jy~km~s$^{-1}$ (after correcting for primary beam attenuation). Thus the total integrated flux observed by FAST is 63.0\% larger than that of observed by THINGS. 
In the area overlapping with THINGS observation, 24.2\% of the fluxes from the single dish measurement is missed by the THINGS map, due to limited short spacing. To further reveal the flux difference, in the top panel of Fig.~\ref{fig:m94_profile} we compare the azimuth averaged surface density of flux profile of FAST and THINGS, and in the bottom panel we compare the accumulated flux profiles. Fig.~\ref{fig:m94_profile} shows clearly that the \ion{H}{1} detected area of FAST is three times that of THINGS and most the excess \ion{H}{1} is distributed in the outskirts of M94 not covered by the THINGS map.

\subsection{Discovery of Discrete Clouds}

Beside the extended disk, we see more diffuse \ion{H}{1} structures outside the disk, which are marked with numbers 1-9 in Fig.~\ref{fig:m94_mom0}.  Cloud 1 is a filament like structure, which could be debris of tidal tails. Cloud 2-7 are discrete clumps distributed around outskirt of M94, which appear to be similar to the high velocity clouds (HVCs) of the Milky Way (MW). Assuming a distance of 4.66~Mpc \citep{K+05},  the projected distance of these clouds is within 50~kpc from the center of M94. 
HVC 8 lies near the center of the disk, so it cannot be seen in the Fig.~\ref{fig:m94_mom0}, but clearly seen in the 382.34~km~s$^{-1}$ panel of the channel map (Fig.\ref{fig:m94_chanmap}), showing a counter-motion with respect to the disk rotation. Remarkably, we also detect a seemingly isolated \ion{H}{1} cloud at a projected distance of 109~kpc south of M94 and name it as Cloud 9, which is much farther from M94 than the MW's respective HVC population. On the FAST high-sensitivity image, we found no any signs of connection between this cloud and the M94 disk, at the column density level of $2.5\times10^{18}$~cm$^{-2}$ (3~$\sigma$). Thus this cloud could be  an isolated object. Indeed, clouds without relatively massive galaxies within a radius of 100~kpc are usually considered as isolated ones \citep{T+17}.  The details about Cloud 9 will be discussed in Section~\ref{section:4.2}. Searching the DELCals images, we found that all these \ion{H}{1} features have no optical counterparts. Table~\ref{table:gas_properties} lists the properties of these \ion{H}{1} features. We use circles to approximate the size of the clouds, with the diameter of these circles listed in column (6). We use an ellipse to approximate \ion{H}{1} filament 1, with its major and minor axis listed in column (6), respectively. Flux of them are integrated within the extent of these ellipse and circles. 

The channel map (Fig.~\ref{fig:m94_chanmap}) focuses on the sky region of $191.89^{\circ}$$\textless$$\alpha$$\textless$$193.50^{\circ}$, $40.80^{\circ}$$\textless$$\delta$$\textless$$41.72^{\circ}$. It shows clear details of the \ion{H}{1} filament 1 and HVC 2-8. Although possible detections HVC 6,8 cannot be distinguished due to superposition in the Fig.~\ref{fig:m94_mom0}, we can clearly see them in the channel map (Fig.~\ref{fig:m94_chanmap}). Moreover, in the 323.83~km~s$^{-1}$ panel of Fig.~\ref{fig:m94_chanmap}, there are two discrete clouds which seems to be connected with the \ion{H}{1} filament and could be the substructures fragmented out of it, hence  we  mark them as 1a and 1b. 

The top panel of Fig.~\ref{fig:m94_369} shows the integrated flux density contours superimposed on the DESI-LS optical image for the HVC 3, 6 and Cloud 9 which are integrated over the velocity range listed in Table~\ref{table:gas_properties}. The FAST contours levels start at 3~$\sigma$ ($8.1\times10^{17}$, $6.8\times10^{17}$, $6.7\times10^{17}$~cm$^{-2}$ for HVC 3, 6 and Cloud 9 respectively)  in steps of 3~$\sigma$. The bottom panels show the spectra toward the peak column density of these clouds. The peak value of each clouds’ spectra exceed 6~$\sigma$. For Cloud 9 the S/N is as high as 9, and the profile of this cloud is the narrowest among the nine features. 

\begin{deluxetable*}{cccccccc}
\tablenum{1}
\tablecaption{Properties of the \ion{H}{1} filament and eight \ion{H}{1} structures\label{table:gas_properties}}
\tablewidth{0pt}
\tablehead{
\colhead{} & \colhead{R.A.} & \colhead{DEC} & \colhead{Velocity Range} & \colhead{Mean Velocity} & \colhead{Diameter} & \colhead{\ion{H}{1} Flux} & \colhead{M$\rm{_{HI}}$} \\
\colhead{...} & \colhead{hh:mm:ss} & \colhead{dd:mm:ss} & \colhead{km s$^{-1}$} & \colhead{km s$^{-1}$} & \colhead{arcmin} & \colhead{Jy km s$^{-1}$} & \colhead{$\times10^{5} M_{\odot}$}
}
\decimalcolnumbers
\startdata
1 & 12:51:42$\sim$12:53:12 & +40:54:30$\sim$+41:17:33 & 260-426 & 355 & 11,26 & 5.646$\pm$0.1 & 288.9$\pm$5.0 \\
2 & 12:51:52 & +41:17:27 & 338-387 & 341 & 4 & 0.125$\pm$0.005 & 6.4$\pm$2.0 \\
3 & 12:51:14 & +41:35:28 & 226-260 & 248 & 4 & 0.100$\pm$0.03 & 5.1$\pm$1.0 \\
4 & 12:50:24 & +41:31:37 & 304-353 & 325 & 6 & 0.329$\pm$0.003 & 16.8$\pm$1.0 \\
5 & 12:50:16 & +41:19:22 & 231-314 & 292 & 8 & 0.714$\pm$0.04 & 36.5$\pm$2.0 \\
6 & 12:48:35 & +41:18:14 & 348-377 & 368 & 4 & 0.080$\pm$0.04 & 4.1$\pm$2.0 \\
7 & 12:48:35 & +41:09:12 & 245-294 & 285 & 8 & 0.683$\pm$0.03 & 34.9$\pm$2.0 \\
8 & 12:51:20 & +41:08:29 & 363-416 & 392 & 6 & 0.980$\pm$0.1 & 50.1$\pm$3.0 \\
9 & 12:51:52 & +40:17:29 & 290-319 & 298 & 6 & 0.239$\pm$0.01 & 12.2$\pm$0.5 \\
\enddata
\tablecomments{
Column 2 and 3: R.A. and DEC (J2000.0) are the Right Ascension and Declination the peak column density of each cloud respectively, expect \ion{H}{1} filament, which is the position range. Column 4: Velocity range is the range that the \ion{H}{1} filament or clouds are visible. Column 5: Mean velocity of HI structures. Column 6: Diameter refers to the diameter of circles similar in size to clouds (major and minor axis respectively for an ellipse of \ion{H}{1} filament 1). Column 7: \ion{H}{1} flux is integrated within the extent of these ellipse and circles. Column 8: M$\rm{_{HI}}$ is the \ion{H}{1} mass of the cloud, in solar masses, calculated by $M\rm{_{HI}}/M_{\odot}$=2.356$\times10^{5}D^{2}S$, where $D$ is the distance to the object in Mpc and $S$ is the \ion{H}{1} flux.
}
\end{deluxetable*}

\subsection{\ion{H}{1} Kinematics}

The velocity field (moment-1) in the same region as the channel map is shown in Fig.~\ref{fig:m94_mom1}. Unlike other \ion{H}{1} clouds, the \ion{H}{1} filament 1 shows a velocity gradient. It is visible from 260.4 to 426.2~km~s$^{-1}$ in the form of a velocity gradient decreasing from the southwest to the northeast. The velocity field between the substructures 1a and 1b of the \ion{H}{1} filament 1 are unchanged. Moreover, HVC 2 and 5 are also discontinuous with the H I disk in velocity, as marked by two black ellipses in Fig.~\ref{fig:m94_mom1}. In addition, we also made moment-2 map \footnote{Note that the spectral resolution of 4.88~km~s$^{-1}$ has not been corrected from this map. Thus it can not show any features with velocity dispersion smaller than 4.88~km~s$^{-1}$.} in Fig.~\ref{fig:m94_mom2}, which indicates the velocity dispersion. Moment-1 and moment-2 map were also made using the spectral-cube algorithm and smoothed by a Guassian filter (kernel=3$\times$3). They were masked at 3 sigma of column density ($2.5\times10^{18}$~cm$^{-2}$ for M94 region, and $6.7\times10^{17}$~cm$^{-2}$ for the lower right subpanel) and a few obvious bad pixels were masked manually.

\begin{figure*}
\begin{center}
	\includegraphics[width=0.9\textwidth]{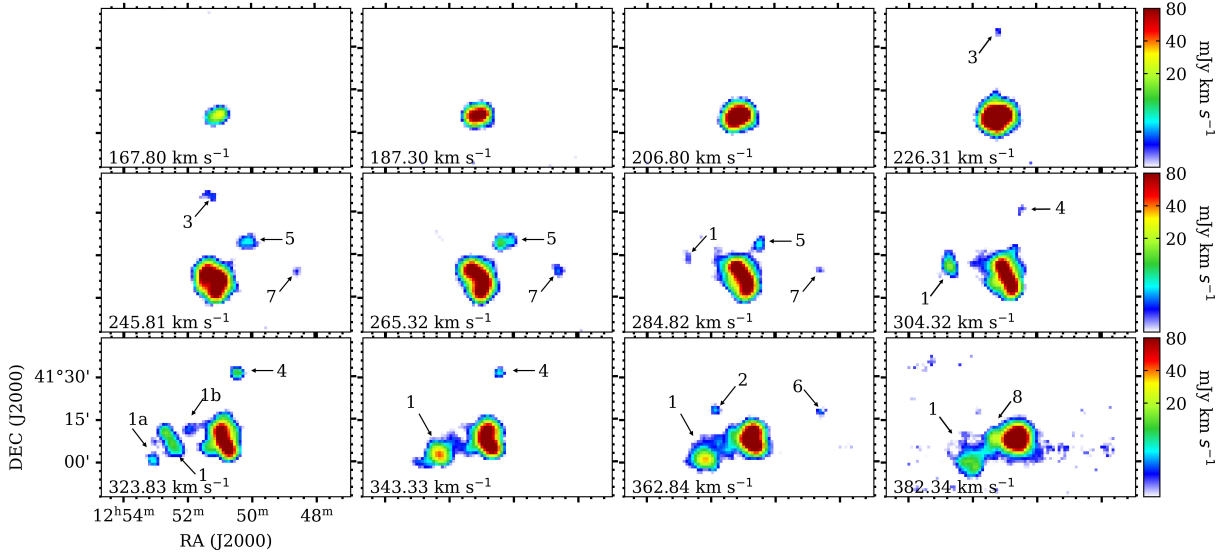}
    \caption{The FAST \ion{H}{1} channel map of the M94 region. Each channel is integrated over a velocity range of 19.5 km s$^{-1}$. Arrows and numbers mark the names of \ion{H}{1} features.
    }
    \label{fig:m94_chanmap}
\end{center}
\end{figure*}

\begin{figure*}
\begin{center}
	\includegraphics[width=0.7\textwidth]{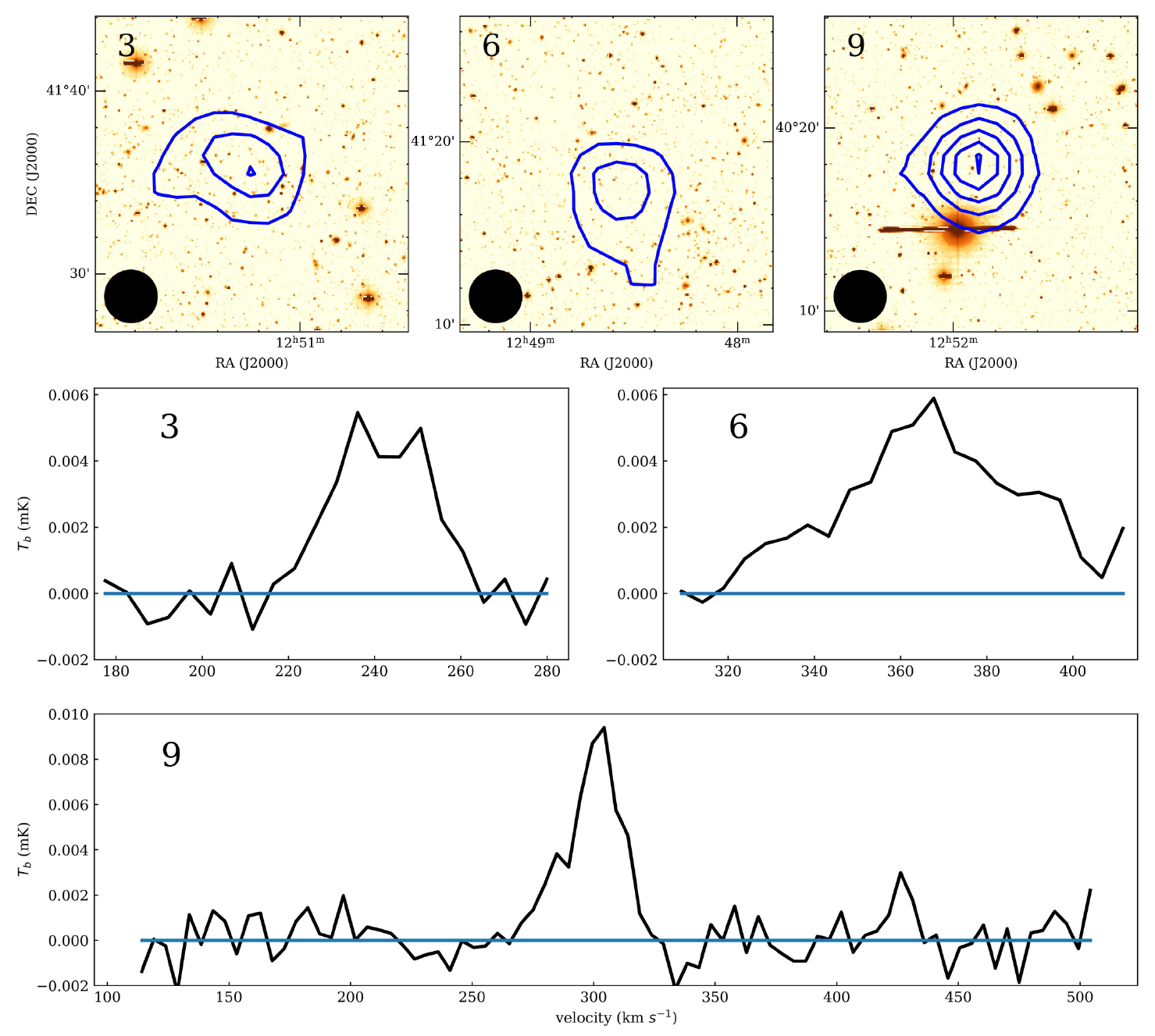}
    \caption{
    {\it Top}: the FAST integrated flux density contours of the \ion{H}{1} detected by FAST superimposed on the DESI-LS optical image for the HVC 3,6 and Cloud 9 respectively, and the integrated velocity range is from Table~\ref{table:gas_properties}. The contour levels start at 3~$\sigma$ ($8.1\times10^{17}$, $6.8\times10^{17}$, $6.7\times10^{17}$~cm$^{-2}$ for HVC 3,6 and Cloud 9 respectively) in steps of 3~$\sigma$. FAST’s HPBW is indicated in the bottom-left corner. {\it Bottom}: spectra taken at the peak column density of the HVC 3,6 and Cloud 9. The vertical axis is 21 cm brightness temperature and the horizontal axis is velocity.
    }
\label{fig:m94_369}
\end{center}
\end{figure*}

To explore the dynamic relationship of the \ion{H}{1} filament and HVC 2,5,8 with M94, we made position-velocity (PV) diagrams along different directions, which are shown in Fig.~\ref{fig:m94_pv}. Panel-a of Fig.~\ref{fig:m94_pv} shows the apparent velocity gradient of \ion{H}{1} filament. From panel-d, we can see that the southwest of \ion{H}{1} filament is not continuous with the \ion{H}{1} disk in velocity which can also be seen in Fig.~\ref{fig:m94_mom1}, although there is a gas connection between the \ion{H}{1} filament and \ion{H}{1} disk in the 343.33 to 382.34~km~s$^{-1}$ panels in Fig.~\ref{fig:m94_chanmap}. Panel-b,-c,-e show that the velocity between HVC 2,5,8 and the \ion{H}{1} disk is discontinuous. As for HVC 3, 4, 6, and 7, they have similar projected distance to the \ion{H}{1} disk, thus they could be remnants of an outer \ion{H}{1} ring. The neutral gas connecting these clumps has disappeared or become undetectable now due to several reasons: either the gas disperses too faintly, or they are ionized by the inter-galactic UV radiation\citep{T+16}. The existence of substructures 1a and 1b indicates that the  \ion{H}{1} filament 1 might start to fragment and can eventually transform to discrete clumps due to dispersion or ionization as well. 

\section{Discussion} 

Compared with the THINGS observations\citep{de+08,W+08}, we detected a more extended \ion{H}{1} disk and found a \ion{H}{1} filament and seven HVCs within 50~kpc of the projection distance from the galactic center. We also found a seemingly isolated \ion{H}{1} cloud located at the projection distance of 109~kpc from the galactic center, which may belong to M94. Here we consider the possible origins of these features.
  
\subsection{Debris of Tidal Features in a Merger?}
\label{section:4.1}

\begin{figure*}
\begin{center}
	\includegraphics[width=0.7\textwidth]{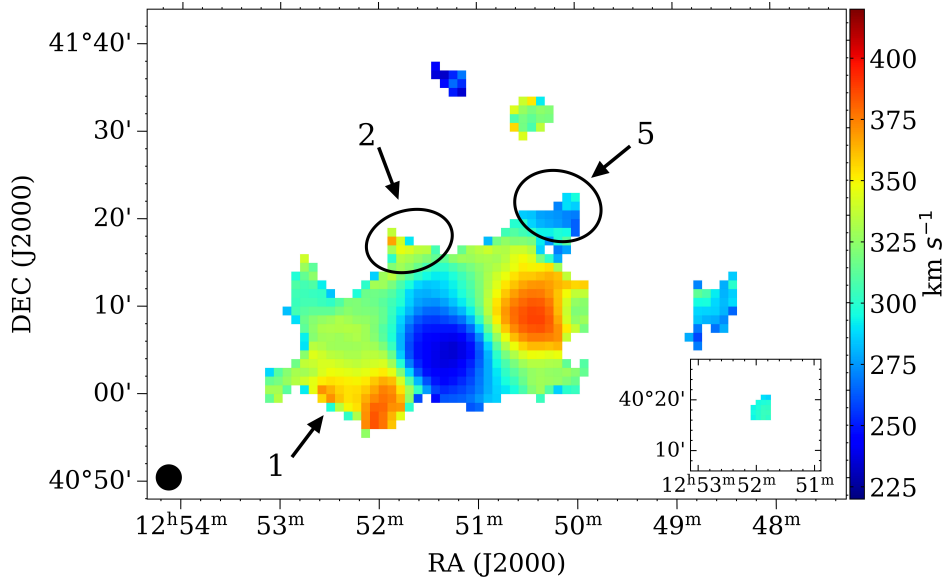}
    \caption{    
     Velocity field of the M94 obtained from the \ion{H}{1} data cube (moment-1 map). The black ellipses represent the HVC 2 and 5. FAST’s HPBW is indicated in the bottom-left corner. The lower right corner shows Cloud 9.
}
\label{fig:m94_mom1}
\end{center}
\end{figure*}

Fig.~\ref{fig:m94_mom0} shows that the \ion{H}{1}  filament 1 appears to follow the spiral arm direction, like most tidal tails such as in the case of M51 \citep{R+90}, NGC 2535 \citep{K+97}, NGC 3893 \citep{V+01}, and NGC 262 \citep{S+87}. Moreover, the velocity field (Fig.~\ref{fig:m94_mom1}) is irregular and disturbed at the position of the \ion{H}{1} filament and HVC 2, 5. Such \ion{H}{1} features without optical counterparts are often found surrounding the interacting systems, for example, NGC 7252 \citep{H+94}.

M94 has a relatively isolated environment, however. It is located near the center of the Canes Venatici I cloud which is a very loose extended system mainly including dwarf irregular galaxies. As the brightest galaxy of the CVn I cloud, M94 is the main disturber relating to only 10 neighboring galaxies \citep{K+05}). Most of the galaxies in this group seem to move with the expansion of the universe, which will prevent nearby collisions from affecting M94 \citep{K+03}. So it seems unlikely for M94 to have any interaction with another members of the group recently.

As for the tidal encounters, \citet{K+03} found no companions with a central surface brightness brighter than 25 $^m/\sq\arcsec$ in the B band within a radius of $\sim$3 degrees or 230~kpc around M94. Although two satellites are found within 100~kpc of its projected distance, the $M_{*}$ of them (9.7$\times$10$^{5}$M$_\sun$ and 6.7$\times$10$^{5}$M$_\sun$) \citep{S+18} are too low to drag these \ion{H}{1} clouds out of the host galaxy. Moreover, there is a LSB galaxy LVJ 1243 +4127 \citep{K+20} at a distance of 4.75$\pm$0.23~Mpc within 120~kpc projected distance of M94, whose mass is also too small to have a major impact. This leads to the idea that M94 may have undergone a past merger. 

The global and large-scale morphology and kinematics of the diffuse gas outside the M94 appears similar to that of the M51. The \ion{H}{1} features in the M51 system is thought to be the results of a very recent ($\sim$~250~Myr) galactic interaction \citep{Y+23}. 
While M94 is seen as a single galaxy, with a giant stellar ring outside the optical disk \citep{T+09} and a corresponding \ion{H}{1} ring \citep{W+08}, could it be a remnant of the galaxy merger that is happening in M51? 

\begin{figure*}
\begin{center}
	\includegraphics[width=0.7\textwidth]{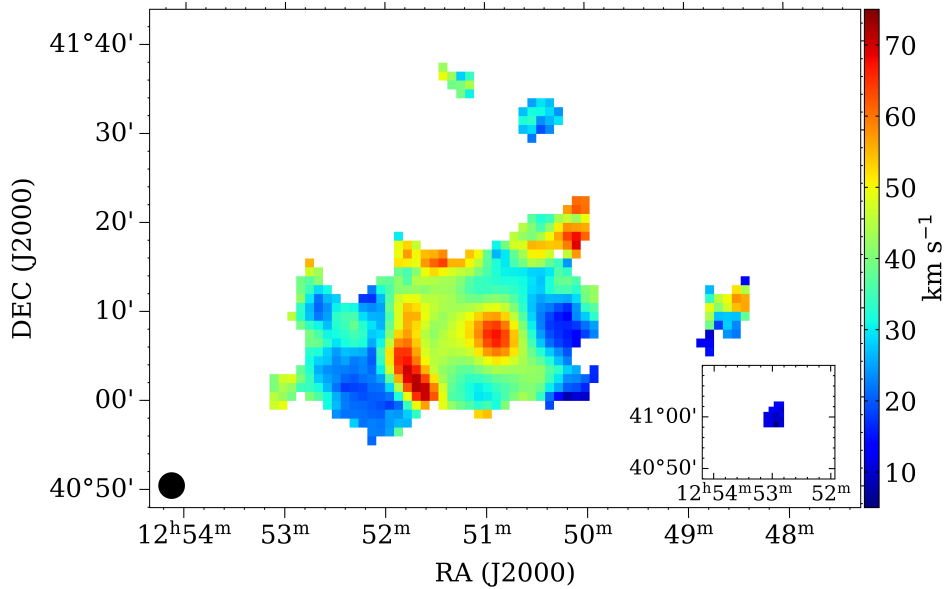}
    \caption{    
     Velocity dispersion of the M94 obtained from the \ion{H}{1} data cube (moment-2 map). FAST’s HPBW is indicated in the bottom-left corner. The lower right corner shows Cloud 9.
}
\label{fig:m94_mom2}
\end{center}
\end{figure*}

To verify such an idea, we turned back to the same simulation data-base as for the M51 system, i.e., the simulations carried out by \citet{Sa+18} \footnote{They have investigated a grid of representative simulations of gas-rich major mergers with mass ratio of 3:1. The total initial baryon mass of the large galaxy is $5.3\times 10^{10} $~M$_{\odot}$ (52\% in gas) and $1.76\times 10^{10} $~M$_{\odot}$ (72\% in gas) for the small galaxy. Both galaxies are embedded in a dark matter halo respectively, assuming a dark-to-baryon mass ratio of 4. The two progenitor disc galaxies are set on a parabolic orbit with a pericentre of 16 kpc.}. Specifically, we compare the period after the fusion of the two progenitor's cores. However, we notice that although the simulation for the M51 system could explain the disk and the ring of M94, it is unable to explain the kinematics of the filament, i.e., the \ion{H}{1} filament 1. The simulation that could explain the observed properties of M94 \ion{H}{1} gas is another model: ``POLAR-RETPRO",which means both galaxies have their disc highly inclined (71 degree) to the orbital plane (``POLAR") and the large galaxy has a RETrograde disc rotation comparing to the angular momentum of the orbit while the second galaxy has a PROgrade rotation. The differences between the M51 model and the M94 model are the disk spins of progenitors with respect to the orbital angular momentum. It is indeed well-known that a polar type orbit favors the formation of ring, \citep[e.g.,][]{H+18}

Fig.~\ref{fig:m94_model} shows the properties, i.e., the \ion{H}{1} column density map and its kinematical maps, of the best-match epoch of the simulation, which is captured at 2 Gyr after the coalescence of the two progenitor's cores. All these maps are created with the same algorithms described in \citet{Y+22}. Specifically, here we adopted a grid resolution of 4~kpc in space and 5~km~s$^{-1}$ in velocity when creating the simulated ``\ion{H}{1}" data-cube. The kinematics maps are obtained by flux-weighted mean in velocity and in dispersion, respectively.

The most peculiar and puzzling feature in observation is the kinematics of the filament (i.e., \ion{H}{1} filament 1) which shows completely different motions as the disk rotation of M94 (Fig.~\ref{fig:m94_mom1}). However, this becomes easy to understand if M94 was formed by a merger of two galaxies as an analog of the model ``POLAR-RETPRO", because the filament could be the tidal tail that was ejected at the first passage of the merger approximately 5 Gyr ago. Since then, the tidal tail continued to expand, with its near part (closer to M94) falling back and feeding the central region of M94.
Due to the projection effect and also the initial geometric properties of the progenitors, i.e., the inclinations with respect to the orbital angular momentum, the tidal tail locates highly inclined (about 45 degrees) to the central disk, as can be seen in Fig.~\ref{fig:m94_model} when comparing the panel-a and -b, for difference views of projection. 
This model could explain why the \ion{H}{1} filament 1 shows a very distinct velocity distribution compared to the central disk rotation. If the filament is the tidal tail in the model, it would be located above the M94 disk and on the near side to us. As a result, we would observe the tidal tail falling towards the central disk, resulting in a velocity gradient from blue to red-shift.

The idea that the filament (\ion{H}{1} filament 1) could be a tidal tail instead being part of the extension of the central disk is supported further by the very high velocity dispersion region (Fig.~\ref{fig:m94_mom2}), which could be also explained by the simulation. The high velocity dispersion is caused by the superposition of the tidal tail and the outer part of the disk: their large contrast motions lead to high values when we calculate the dispersion in velocity. 
The infalling process of a tidal tail could last for several billion years or even longer, provided the tidal tail can survive the cosmic UV background radiation, for example.

If the \ion{H}{1} filament and HVCs were formed simultaneously by the first passage of the merger 5 Gyr ago, they may have undergone ionization as a result of photon ionization by young stars and the comic background. \citet{Mo+10} estimated the dimensionless ionization parameter \citep{T+11} of M94, which is log(U)~$\sim$~-3.04~$\pm$~0.24. To estimate the neutral ratio of hydrogen (N$\rm{_{HI}}$/N$\rm{_{H}}$) of these \ion{H}{1} features after 5~Gyr of photo ionization, we take HVC 6 as an example,  which has the lowest column density (2.04$\times10^{18}$~cm$^{-2}$). We fit the values of logU and N$\rm{_{HI}}$ of HVC 6 in the model in Figure 24 of \citet{W+23}, and found that the resulting lower limit of N$\rm{_{HI}}$ is 1.8\%N$\rm{_{H}}$, which indicates 5~Gyr of photo ionization does not ionize all \ion{H}{1}. Another effect needs to be considered is  the effect of evaporation. Using Equation (10) and similar parameters from \citet{T+16}, we estimated that only clouds with a radius of at least 26.8~kpc can survive after 5~Gyr evaporation time given the column density of the \ion{H}{1} filament (7.91$\times10^{18}$~cm$^{-2}$). Thus, these HVCs should be remnant fragments of larger structures rather than isolated clouds initially. There might be more than one tidal tails ejected at the first passage of the merger 5~Gyr ago, which dispersed into smaller clouds. We also estimate that after another 1.4~Gyr evaporation time, the \ion{H}{1} filament 1 will disperse into smaller clouds and HVC 2-8 will become too faint to be detected.

\subsection{Possible Origins of Cloud 9}
\label{section:4.2}

\begin{figure*}
\begin{center}
	\includegraphics[width=0.9\textwidth]{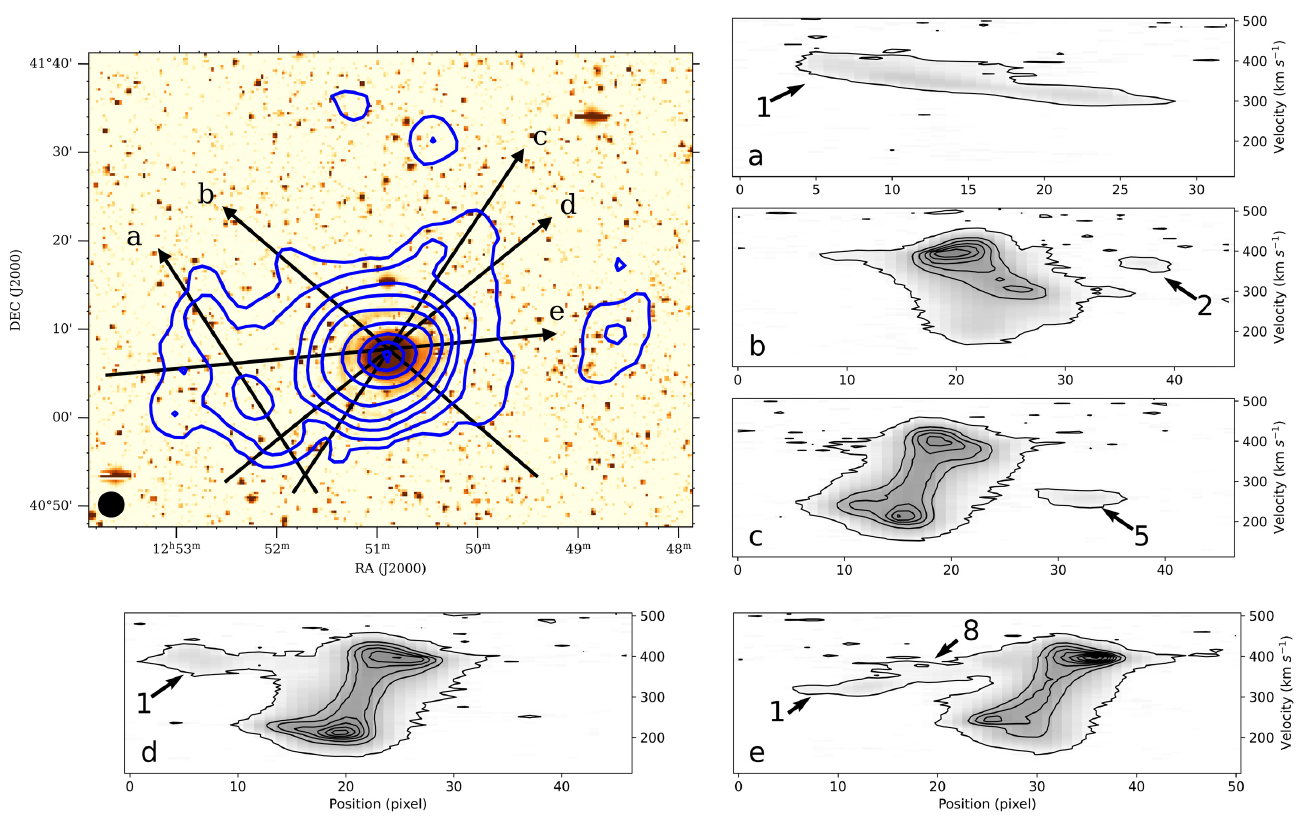}
    \caption{
    Four position-velocity plots along different directions. {\it Top-left}: the FAST \ion{H}{1} integrated flux density contours in blue color overlaid on the DESI-LS optical image. The contour values are the same as those in Fig.~\ref{fig:m94_mom0}. The five black arrows mark the directions and positions of position-velocity diagrams in panels a-e, respectively. FAST’s HPBW is indicated in the bottom-left corner. {\it Panels a-e}: the black contours start at 4~$\sigma$ (16.3 mJy) in steps of 80~$\sigma$. Arrows and numbers mark the names of \ion{H}{1} features.
    }
    \label{fig:m94_pv}
\end{center}
\end{figure*}

\begin{figure*}
\begin{center}
	\includegraphics[width=0.9\textwidth]{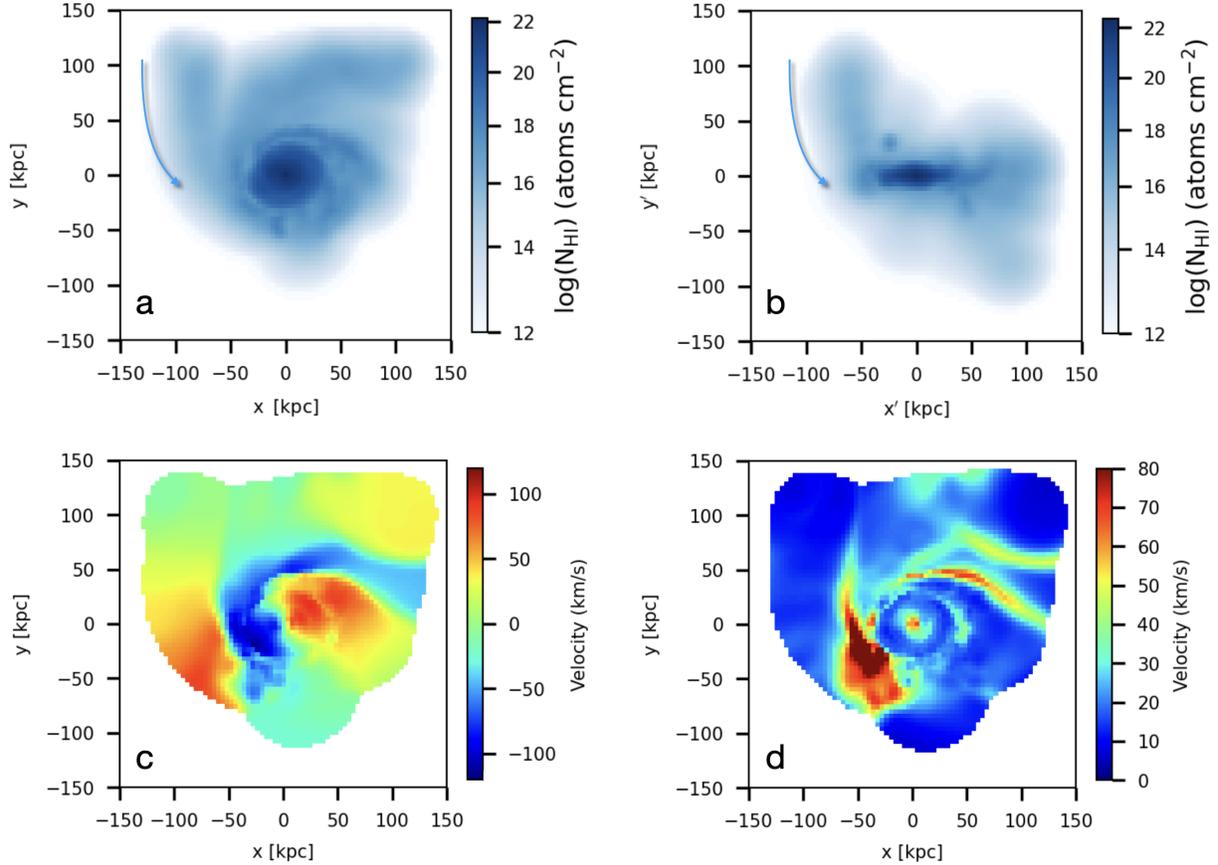}
    \caption{
    Morphology and kinematics of the simulated galaxy that is captured from the major merger model at 2 Gyr after the fusion of the two progenitors (see text for more details). The \ion{H}{1} column density (panel-a), the velocity field (panel-c) and velocity dispersion map (panel-d), projected at the observational view. Panel-b, the \ion{H}{1} column density map projected along the edge-on view of the gas disk. The blue arrows in Panel-a and -b  indicate the moving directions of the tidal tail from the secondary progenitor, respectively, in each projected plane.
    }
    \label{fig:m94_model}
\end{center}
\end{figure*}

Cloud 9 is seemingly isolated which may belong to M94, and its projection distance from the galactic center is 109~kpc. It’s \ion{H}{1} flux is 0.14$\pm$0.02 Jy km s$^{-1}$, corresponding to a mass of (7.2$\pm$1.0)$\times10^{5}~M_{\odot}$. And the velocity width W$\rm{_{50}}$ of Cloud 9 is about 20 km s$^{-1}$. Adopting the Equation (2) from \citet{Wo+16}, we take r$\rm{_{1/2}}$=3.49~kpc, FWHM=20~km~s$^{-1}$, and we estimate the dynamical (or viral) mass M$\rm{_{dyn}}$ of Cloud 9 as 3.49$\times10^{8}~M_{\odot}$. Here we consider several possible origins of Cloud 9 specially. 

Firstly, Cloud 9 could be tidal debris composed of pure gas without dark matter. In this case, the line width broadening is dominated by the gas temperature and is given by T/K=21.8 $\rm{{W^{2}_{50}}}$ with W$\rm{_{50}}$ given in km s$^{-1}$. Using W$\rm{_{50}}$=20~km~s$^{-1}$ yields a gas temperature T$\sim$8.7$\times10^{3}$K. Considering the distance between Cloud 9 and M94, it is not unusual to find tidal features longer than 100~kpc in merging systems \citep[e.g.][]{A+87,S+12,K+07}. Thus it is possible to find tidal debris far away from the parent galaxies after merger. Based on the NGC 4490/85 system, \citet{P+18} present a detailed theoretical model of the tidal encounter between two isolated low-mass galaxies. They demonstrated that baryons can be ‘parked’ at very large distances by repeated encounters between two dwarf galaxies. If Cloud 9 is originated from the same passage of the merger as other \ion{H}{1} features, the lower limit of the initial diameter of it should be 26.8~kpc according to the results of Section~\ref{section:4.1}.  Known \ion{H}{1} streams could persist for several hundreds of Myr to a Gyr or more, but few \ion{H}{1} stream can survive as long as 5~Gyr. For example, \citet{O+05} suggest a 100~kpc plume in Virgo has persisted for $\textgreater$100~Myr; the model of \citet{Mi+10} implies the formation of the Leo Ring has begun 1.2~Gyr ago. If we consider the possibility that Cloud 9 is tidal debris which is the remnants of a long tidal tail, 5~Gyr seems to be too long for it to survive. On the other hand, mechanisms involve  star formation or dark matter or both of them seem to be more ready to explain the long survival time of Cloud 9, as discussed in the following.

Secondly, Cloud 9 may be a dark dwarf galaxy, such as a low surface brightness (LSB) galaxy. The optical detection limits of DESI-LS is 29.15 mag~arcsec$^{-2}$ in g, 28.73 mag~arcsec$^{-2}$ in r, and 27.52 mag~arcsec$^{-2}$ in z \citep{M+23}. Adopting the method from \citet{Z+21}, we estimated that Cloud 9 would not be brighter than 25.66 mag in g, 25.24 mag in r, and 24.03 mag in z, with the disk scale length R$\rm{_{s}}$=0.080~kpc, the major-to-minor axis ratio q=1. So Cloud 9 would not be brighter than 26.03 mag in B-band as estimated with the equation $m\rm{_{B}}=m\rm{_{g}}+0.47(m\rm{_{g}}$$-m\rm{_{r}})+0.17$ \citep{S+02}. We can further derive that the lower limit absolute B-band magnitude (M$\rm{_{B}}$) is 2.31 mag, and the upper limit B-band luminosity is 1.26$\times10^{3}~L_{\odot}$ obtained with $L\rm{_{B}}=D^{2}10^{10-0.4(m\rm{_{B}}-M\rm{_{B0}})}$, where M$\rm{_{B0}}$ is the absolute solar B-band magnitude adopted as 5.44 mag \citep{W+18}. Moreover, we estimated the upper limit stellar mass of Cloud 9 is 1.14$\times10^{5}~M_{\odot}$, with the g-r color value and B-band luminosity \citep{Z+17}. We can calculate the baryonic mass (M$\rm{_{bar}}$) of the probable dark galaxy associated with Cloud 9 by $M\rm{_{bar}}=M\rm{_{*}}+M\rm{_{gas}}$, where M$\rm{_{*}}$ is the stellar mass and M$\rm{_{gas}}$ is the total gas mass. We can ignore the contribution of the stellar mass to the baryonic mass of Cloud 9, because it is too small. Assuming the same helium-to-\ion{H}{1} ratio as that obtained from Big Bang nucleosynthesis, the total gas mass (M$\rm{_{gas}}$) is determined by $M\rm{_{gas}}=1.33\times M\rm{_{HI}}$. Thus the baryonic mass of Cloud 9 is about 9.58$\times10^{5}~M_{\odot}$. The dynamical mass to total baryonic mass ratio is derived to be about 364, implying that dark matter absolutely dominates over baryons in Cloud 9. We used the Baryonic Tully-Fisher relation (BTF) to explore the properties of the probable dark galaxy Cloud 9 which is an empirical correlations for disk galaxies. This dark galaxy is gas-dominated, but it nearly follows the best-fit relation $M_{bar}=A{V^{4}_{rot}}$ with $A=47\pm6$~$M_{\odot}km^{-4}s^{4}$ for gas-dominated disk galaxies \citep{M+12} and $V\rm{_{rot}}=1/2$ $\rm{FWHM}$. This implies that Cloud 9 may be hosting a normal dwarf disk galaxy. 

Thirdly, Cloud 9 may also be a mini dark matter halo only composed of gas and dark matter, and hence it is invisible in optical. Supposing it is in Virial equilibrium, it is sensible to equal M$\rm{_{200}}$ \footnote{We define M$\rm{_{200}}$ as the mass calculated within a radius with mean inner density equals 200 times the critical density of the Universe, $\rho_{crit}=3H^{2}/8\pi G$.} to M$\rm{_{dyn}}$. Cloud 9 may be a starless gaseous minihalo of present-day mass in the range $10^{6}$$\lesssim$M$\rm{_{200}}$/$M_{\odot}$$\lesssim$5$\times10^{9}$, in which the gas is in thermal equilibrium with the UV background radiation and in hydrostatic equilibrium in the gravitational potential of the halo, i.e. a RELHIC object \citep{A+17,A+20}. Cloud 9 conforms to several properties of RELHICs in the Local Group proposed by \citet{A+17}: (i) nearly round on the sky (b/a$\textgreater$0.8 at 1$\times10^{18}$~cm$^{-2}$) (ii) with a small (sub-kpc) neutral hydrogen core (iii) a very narrow distribution of thermally broadened line widths (W$\rm{_{50}}\sim$20~km s$^{-1}$). If Cloud 9 was a RELHIC, it is provide a robust prediction of the CDM paradigm and a great example to explore the RELHICs outside the Local Group.

Comparing the above three origins of Cloud 9, we prefer the dark dwarf galaxy and RELHIC scenarios rather than tidal debris based on the estimated long survive time of this cloud.

\section{Conclusion}

In conclusion, we have obtained a deep \ion{H}{1} image for the vicinity of M94 with FAST and discovered much more diffuse \ion{H}{1} structures. Major new features are listed in the following:

\begin{itemize}[leftmargin=*]
\item A more extended \ion{H}{1} disk.
\item A \ion{H}{1} filament and seven HVCs, whose projected distance from the galactic center is within 50~kpc.
\item A seemingly isolated Cloud 9 at the projection distance of 109~kpc.
\end{itemize}
With the complex morphology and kinematics, we suggest that M94 is likely a remnant of a major merger of two galaxies. To explain the observed properties of M94 \ion{H}{1} gas, we simulated it with the model: ``POLAR-RETPRO”. The best-match epoch of the simulation is captured at 2~Gyr after the coalescence of the two progenitor’s cores. We suggest that the \ion{H}{1} filament could be the tidal tail ejected at the first passage of the merger 5~Gyr ago. The HVCs could be fragments of larger structures such as tidal tails originated at the same time as the \ion{H}{1} filament. As for the nearly isolated Cloud 9, we consider several possible origins, especially the dark dwarf galaxy and RELHIC scenarios. The former is expected to contribute considerably to our understanding of the low-mass and ultra-faint galaxies. And the detection and characterization of the latter would provide a novel opportunity to reveal the “dark” side of a CDM-dominated universe. In optical images, M94 is such an isolated galaxy with few signs of interaction. Thus high-sensitivity \ion{H}{1} imaging is important in revealing the diffuse HI structures and tidal debris which is crucial to understanding the dynamical evolution of galaxies.

\begin{acknowledgments}
We thank the FAST staff for help with the FAST observations. We thank Dr. Francois Hammer for his helpful discussion. 
We acknowledge the support of the National Key R\&D Program of China (2018YFE0202900; 2017YFA0402600).
We also thanks the National Natural Foundation of China (NSFC No. 12041302 and No. 11973042), also the support the International Research Program (IRP) Tianguan, which is an agreement between the CNRS, NAOC and the Yunnan University.
\end{acknowledgments}


\end{document}